\documentclass[numsec,webpdf,modern,medium]{oup-authoring-template}

\onecolumn 

\setcitestyle{authoryear,round} 

\graphicspath{{Fig/}}

\usepackage{booktabs}
\usepackage{makecell}
\usepackage{rotating}
\theoremstyle{thmstyleone}%
\theoremstyle{thmstyletwo}%
\theoremstyle{thmstylethree}%

 \usepackage{hyperref}
 \usepackage{multirow}
\usepackage[table]{xcolor}
\usepackage{tabularx}
\usepackage{adjustbox}
\usepackage{tikz}
\usetikzlibrary{arrows.meta,positioning,shapes.geometric,backgrounds,fit}
\usepackage{enumerate}

\usepackage{etoolbox}

\makeatletter

\apptocmd{\ps@opening}{%
  \let\@oddhead\@empty
  \let\@evenhead\@empty
  \let\@oddfoot\@empty
  \let\@evenfoot\@empty
}{}{%
  \PackageWarning{header-fix}{Could not patch opening page style}%
}

\apptocmd{\ps@headings}{%
  \def\@evenhead{%
    \vbox{%
      \hbox to \textwidth{%
        {\fontsize{6.5bp}{6.5bp}\selectfont
         \sffamily\bfseries\thepage}%
        \hfill
      }%
      \vspace{5.5\p@}%
      {\color{black!60}\rule{\textwidth}{1\p@}}%
    }%
  }%

  \def\@oddhead{%
    \vbox{%
      \hbox to \textwidth{%
        \hfill
        {\fontsize{6.5bp}{6.5bp}\selectfont
         \sffamily\bfseries\thepage}%
      }%
      \vspace{5.5\p@}%
      {\color{black!60}\rule{\textwidth}{1\p@}}%
    }%
  }%
}{}{%
  \PackageWarning{header-fix}{Could not patch headings page style}%
}

\AtBeginDocument{\pagestyle{headings}}

\makeatother

\begin{document}

\journaltitle{}
\DOI{}
\copyrightyear{}
\pubyear{}
\vol{}
\issue{}
\access{}
\appnotes{}

\firstpage{1}
\title[Short Article Title]{Enhancing Gender Equality Assessment through Object-Oriented Bayesian networks: the European Gender Equality Index Case}
\author[1,$\ast$]{Lorenzo Giammei\ORCID{0009-0009-3976-0107}}
\author[2]{Fulvia Mecatti}
\author[3]{Flaminia Musella}
\author[4]{Paola Vicard}

\address[1]{\orgname{National Institute for Public Policy Analysis}, \orgaddress{\street{Corso d’Italia, 33}, \postcode{00198}, \state{Rome},
\country{Italy}}}
\address[2]{\orgdiv{Department of Sociology and Social Research}, \orgname{University of Milano-Bicocca}, \orgaddress{\street{Via Bicocca degli Arcimboldi, 8}, \postcode{20126}, \state{Milan}, 
\country{Italy}}}
\address[3]{\orgdiv{Education Department}, \orgname{University Roma Tre}, \orgaddress{\street{Via del Castro Pretorio, 20}, \postcode{00185}, \state{Rome}, 
\country{Italy}}}
\address[4]{\orgdiv{Department of Economics}, \orgname{University Roma Tre}, \orgaddress{\street{via Silvio d'Amico, 77}, \postcode{00145}, \state{Rome}, \country{Italy}}}

\corresp[$\ast$]{Corresponding author. \href{email:l.giammei@inapp.gov.it}{l.giammei@inapp.gov.it}}

\received{Date}{0}{Year}
\revised{Date}{0}{Year}
\accepted{Date}{0}{Year}

\abstract{A novel data-driven framework is introduced to assess gender equality by complementing and empowering a widely used European gender composite indicator, the Gender Equality Index (GEI). The GEI synthetizes the latent construct of gender equality into a single score and is extensively employed for cross-country comparison and monitoring. While effective for communication and benchmarking, this practice is affected by conceptual and methodological limitations, including marginal analysis that leaves interactions and conditional (in)dependencies unmeasured, and a lack of predictive capability. 
To address these limitations, this paper proposes the use of Object-Oriented Bayesian Networks (OOBNs) to model the GEI. By preserving the hierarchical structure of the index, OOBNs extend Bayesian Networks and enable a multivariate and probabilistic representation of interdependencies among the components of gender equality. This approach advances intersectional gender statistics by shifting the focus from computing a single composite score to modelling the underlying mechanisms that shape gender inequalities.
The proposed methodology enhances the assessment and monitoring of gender equality and adds a predictive dimension through scenario-based evaluation, thereby supporting Gender Impact Assessment and policy decision-making. An application to Italian official statistics illustrates the practical relevance of the framework and its applicability to other national contexts and policy needs.
}


\maketitle

\section{Introduction}
\label{sec1}

Gender equality is widely recognized as a fundamental human right, a core democratic value, and a cornerstone of social and economic development. Its achievement is a prominent goal of the United Nations (UN)  2030 Agenda for Sustainable Development \citep{UN2015} and a constitutional principle of the European Union (Article 2 and Article 3(3)) of the Treaty on the European Union \citep{EU2012}.
As emphasized by the UN, the persistence of gender inequality affects half of the world’s population and constrains the potential of society as a whole. Reducing gender inequalities in all aspects of life is therefore both a civilisational objective and a recognized driver of social and economic growth.

To pursue these aims, gender statistics play a crucial role by providing the empirical evidence necessary to inform data-driven policies, allocate resources effectively, raise public awareness, and support change.
Among existing instruments, the Gender Equality Index (GEI), released annually  by the European Institute for Gender Equality (EIGE) since 2010, has become a key reference for providing a valuable benchmark of gender equality across the European Union.  

Nevertheless, as a composite indicator \citep{Saltelli2020, OECD2008}, the GEI inevitably reduces a complex system of interrelated dimensions, selected to reflect the multi-faceted notion of gender equality, into a single score. This rough aggregation, while useful for comparability, may obscure the underlying structure of relationships, and may limit the scope for intersectional and scenario-based analyses. This poses an ever-present need for improved state-of-the-art statistical tools to tackle gender gaps and to assess gender equality within diverse contexts: national, local, sub-group, or thematic.  

In previous work \citep{mecattibayesian2022}, Bayesian Networks (BNs) were proposed as a means to move beyond the single GEI score by modeling multivariate relationships among its components.

The present paper advances this line of research by introducing Object-Oriented Bayesian Networks (OOBNs) as a methodological framework tailored to the GEI. OOBNs preserve the layered architecture of the GEI (see Section \ref{subsec23}), while overcoming the limitations of traditional composite indicators as well as those of standard BNs. As statistical learning models, OOBNs provide enriched outputs that support more rigorous gender and intersectional analyses, while at the same time endowing the framework with predictive capacity and scenario-building capabilities.

The contribution of this paper is twofold: it develops and illustrates an OOBN model for the GEI, showing how this approach enhances the measurement, monitoring, and assessment of gender equality as well as the scenario-based evaluation of policy impacts; and it applies the model to Italian data, demonstrating how OOBNs empower the GEI with what-if analysis and deliver more informative evidence for policy and decision-making.

The remainder of the paper is organized as follows. 
Section 2 provides background by reviewing the evolution of gender statistics, the role and limitations of composite indicators, and the architecture of the GEI as the reference framework of this study. Section 3 introduces the methodological contribution, showing how OOBNs extend BNs and advance their modeling of the hierarchical structure of the GEI. The Italian case study is presented in Section 4, which addresses country-specific challenges, estimates the OOBN for the Italian GEI at a finer level of granularity, and illustrates its use for scenario analysis and gender impact assessment. Section~5 concludes with a discussion of the strengths and limitations of the proposed methodology, and points to directions for future research.

\section{Measuring Gender Equality: State of the Art and Open Challenges}
\label{sec2}

\noindent This section provides the background for our methodological contribution, with a concise account of the emergence and evolution of gender statistics to contextualize the development of current frameworks for assessing gender equality and to focus on the GEI as the reference framework of the present research.

\noindent For the purposes of this paper, we consider the term \emph{gender} as distinguished from \emph{sex}, according to WHO definitions \citep{WHOgender}. In essence, while sex refers to biological characteristics, gender encompasses socially constructed norms, roles, and relationships, which interact with but differ from biological sex.  

Defining \emph{gender equality} is less straightforward, as it represents a multidimensional concept that entails rights, responsibilities, and opportunities for participation in all domains of life. Far from implying that women and men must become the same, gender equality requires that they enjoy equal conditions to participate fully in society. Enshrined in the 1948 UN Universal Declaration of Human Rights, it has progressively been recognized as both a cornerstone of democracy and a driver of economic and social development. This recognition has generated a growing demand for gender-sensitive data and for statistical analysis explicitly framed under a gender perspective. Gender-based disparities have direct costs, therefore any action aiming at reducing them will potentially benefit the whole society, both women and men. Gender statistics play a crucial role in formulating evidence-based policies and programs to address gender biases and promote gender equality. 

\subsection{Gender Statistics in Brief: Origins and Evolution}
\label{subsec21}

The emergence of gender statistics can be traced through five milestones that illustrate how international agendas have progressively incorporated gender-sensitive data into official monitoring frameworks:
\begin{enumerate}[(a) ]
    \item In 1975, proclaimed the UN International Year of Women, the First World Conference on Women in Mexico marked a turning point from a gender-blind to a female-centered perspective. The subsequent UN Decade for Women (1976–1985) promoted dedicated programmes on gender statistics, launched for instance by the UN Statistical Division \citep{UNDESAStats}.
    \item The 1995 Beijing Platform for Action urged national statistical offices to strengthen gender analysis and ensure that statistics reflect the realities of both women and men. This consolidated the conceptual shift from a women-focused view to a broader perspective on gender relations. 
    \item With the 2000 Millennium Declaration, the eight Millennium Development Goals (MDGs) institutionalized systematic monitoring of gender equality. Goal~3 explicitly targeted gender equality and women’s empowerment, committing countries to systematically monitor and report their progress on the basis of a shared system of measurable parameters and gender-specific indicators. This also emphasized that reducing gender disparities is both a vehicle and an accelerator for achieving all other MDGs. 
    \item In 2007, the first UNSD Global Forum on Gender Statistics was convened (and has convened bi-annually since), establishing a permanent space for producers and users to review methodologies, analytical solutions, and best practices in gender-sensitive data. 
    \item At the 2015 deadline, with several Millennium Development Goals (MDGs) largely unmet, nations were urged to commit to the United Nations 2030 Agenda with its core list of 17 Sustainable Development Goals (SDGs). In this transition, gender equality was reaffirmed as Goal~5, namely to achieve gender equality and empower all women and girls, defined as a necessary foundation for a peaceful, prosperous, and sustainable world \citep{UN2015}. Monitoring, conducted through 14 official indicators covering the 9 targets of SDG~5, has renewed impetus to the steady flow of official gender statistics towards public platforms, e.g. the World Bank gender data portal, UN-Women data hub, and the European Institute for Gender Equality gender statistics database \citep{WorldBankGender, UNWomenDataHub, EIGEgenderstats}.
\end{enumerate}

In parallel with the institutionalization of gender statistics, the Beijing Platform for Action also established \emph{gender mainstreaming} as the global strategy for achieving gender equality, requiring that a gender perspective be systematically integrated into the design, implementation, monitoring, and evaluation of all policies and programmes (for articulated definitions and extensive discussion see \cite{Hannan2022}). Because legislation, regulations, and projects are rarely gender-neutral, disregarding gender differences risks perpetuating or even reinforcing structural inequalities. 

A cornerstone of gender mainstreaming is the \emph{Gender Impact Assessment} (GIA), which provides a step-wise framework for \emph{ex-ante} analysis of planned actions. GIA enables policymakers to anticipate differential effects on women and men, to identify gaps, and to prevent unintended consequences \citep{CouncilEurope2004}. While mainly applied in the early stages of the policy cycle (design and planning), GIA also supports monitoring and evaluation across the implementation process. Moreover, its scope extends beyond the public sector: by aligning with the principles of Responsible Research and Innovation (RRI), GIA is increasingly relevant in contexts of technological and social innovation, where structural disadvantages by gender or other intersecting categories risk being reproduced \citep{Dahlin2023}. For more detailed definitions of GIA and extensive discussion, see among others \cite{EIGE2017}.

In this perspective, recalling gender mainstreaming and GIA serves to highlight the broader policy framework in which our methodological contribution is situated. The use of OOBNs to empower the GEI can be seen as a statistical form of GIA, responding to the need for advanced tools able to capture the complexity of gender equality. We provide a framework that advances intersectional gender statistics analysis and adds predictive capacity through ex-ante evaluation and what-if analysis. In this way, OOBNs extend the reach of gender statistics and empower the GEI to function not only as a monitoring device but also as a decision-support tool for more effective gender-sensitive policies.

\subsection{Gender Composite Indicators}
\label{subsec22}

Gender equality is widely recognized as a complex, multidimensional, latent  construct: it cannot be observed directly, but only through a set of proxy categorisations, counts and measures that reflect its many facets. Moreover, inequalities intersect with other structural divides, such as age, socioeconomic status, ethnicity, disability, or geography. This intersectionality is now a central perspective in gender research and policy \citep{Nedera2023}, and in statistical terms it requires a full multivariate approach in producing and analysing gender statistics.

\begin{figure}[htp]
    \centering
    \includegraphics[width=14cm]{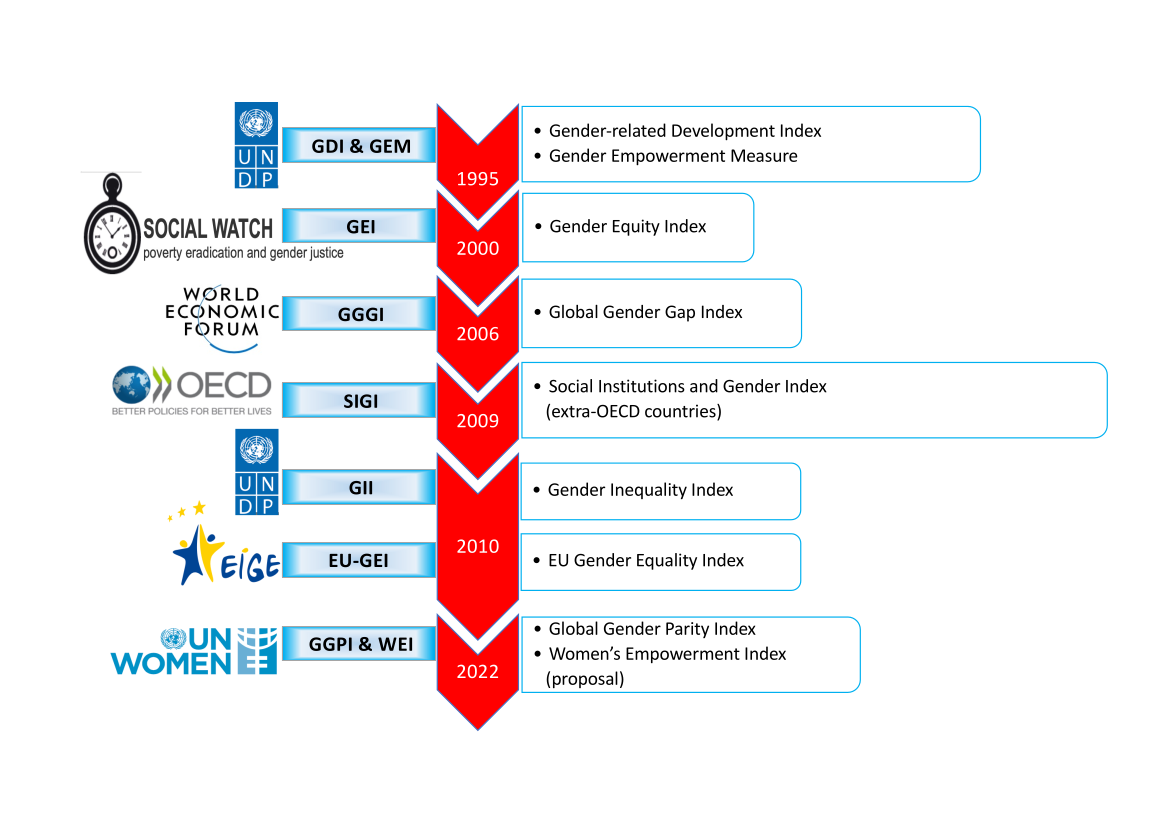}
    \caption{Main gender composite indicators released by international organizations: source, acronym, year of first publication, extended name}
    \label{fig:Fig1}
\end{figure}

A widely adopted practice to monitor and compare progress towards gender equality is the use of composite indicators. Unlike simple indicators (e.g., ratios or averages), composite indicators aggregate sets of simple indicators into a single score. Gender composite indicators aim to capture the multidimensional nature of gender equality in a synthetic and comparable form, providing country rankings and benchmarks of performance. Several have been developed and are periodically disseminated by supra-national organizations (Figure~\ref{fig:Fig1}), with the chronology reflecting both evolving conceptual frameworks and advances in quantification methods (see for instance \cite{Mecatti2012}), from the adaptation of development indicators in the 1990s to more recent proposals of "female capabilities"-based indexes \citep{Berik2022}.

Composite indicators are successful because they compress a complex, latent dimension into an accessible and communicable score, widely used in official statistics, policy analysis, and public debate. Yet, they also suffer from conceptual and methodological drawbacks common to all composite indicators 
 (see among others, \cite{OECD2008}, \cite{Permanyer2010}, \cite{ALAIMO2024101712}). Their construction follows the same step-wise procedure: (i) selecting domains and sub-domains as a qualitative representation of gender equality; (ii) identifying measurable variables and related simple indicators synthesizing observed data (often as averages or women-to-men ratios); (iii) optional data processing such as normalization and treatment of missing values; (iv) aggregating indicators within domains; and (v) aggregating across domains to obtain a single normalized value, usually scaled between 0 and 1 (or 1–100).  

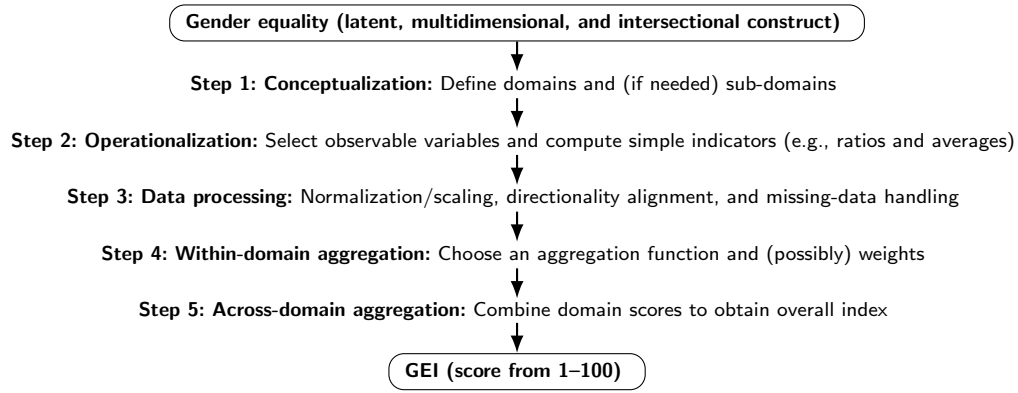
\begin{figure}[htbp]
\centering

\begin{tikzpicture}[
    node distance=4mm,
    >=Latex,
    font=\footnotesize,
    io/.style={
        draw,
        rounded corners=2mm,
        align=center,
        inner xsep=6pt,
        inner ysep=3pt,
        font=\footnotesize\bfseries
    },
    middle/.style={
        draw=none,
        align=center,
        inner sep=1pt
    },
    arrow/.style={
        ->,
        thick
    }
]

\node[io] (start) {
    Gender equality (latent, multidimensional, and intersectional construct)
};

\node[middle, below=of start] (s1) {
    \textbf{Step 1: Conceptualization:}
    Define domains and (if needed) sub-domains
};

\node[middle, below=of s1] (s2) {
    \textbf{Step 2: Operationalization:}
    Select observable variables and compute simple indicators
    (e.g., ratios and averages)
};

\node[middle, below=of s2] (s3) {
    \textbf{Step 3: Data processing:}
    Normalization/scaling, directionality alignment, and missing-data handling
};

\node[middle, below=of s3] (s4) {
    \textbf{Step 4: Within-domain aggregation:}
    Choose an aggregation function and (possibly) weights
};

\node[middle, below=of s4] (s5) {
    \textbf{Step 5: Across-domain aggregation:}
    Combine domain scores to obtain overall index
};

\node[io, below=of s5] (out) {
    GEI (score from 1--100)
};

\draw[arrow] (start) -- (s1);
\draw[arrow] (s1) -- (s2);
\draw[arrow] (s2) -- (s3);
\draw[arrow] (s3) -- (s4);
\draw[arrow] (s4) -- (s5);
\draw[arrow] (s5) -- (out);

\end{tikzpicture}

\caption{Step-wise construction of the GEI as a composite indicator.}
\label{fig:flow_composite_indicator_simple}
\end{figure}

Each step involves choices (as displayed in Figure~\ref{fig:flow_composite_indicator_simple}) that are subjective and debated: which domains best represent gender equality, which variables and indicators to include, whether and how to normalize, how to aggregate (e.g., linear or geometric means), and whether and how to weight aggregations. Different decisions can lead to substantially different scores and country rankings, even when intended to measure the same gender equality/inequality concept. In addition, composite indicators reduce a rich multidimensional reality into one number, inevitably losing information. Despite commonly being referred to as “multivariate indexes,” they in fact aggregate marginal statistics one variable at a time. Interactions and conditional dependencies among variables — where gender intersectionality resides — remain unmeasured. As argued by \cite{Berik2022}, such indicators risk oversimplifying and “shrinking” the very concept they intend to capture.

Nevertheless, gender composite indicators remain valuable as entry points: they provide visibility, comparability, and communication power, and they stimulate debate on gender equality. Their limitations, however, underline the need for advanced statistical tools able to preserve the layered structure of composite indicators, account for interdependencies, and support scenario-based analysis. The methodological contribution presented in the following sections — the use of Object-Oriented Bayesian Networks applied to the GEI — is proposed precisely in this perspective.

\subsection{Focusing on the European Gender Equality Index: Why and How}
\label{subsec23}

Among the several gender composite indicators shown in Figure~\ref{fig:Fig1}, this paper focuses on the European Gender Equality Index (GEI, labelled EU-GEI in the figure, henceforth simply GEI). The choice is motivated by three main considerations. First, the GEI is among the most recent, and thus better reflects conceptual and methodological advancements in measuring gender equality. Second, it is limited to Europe, which improves comparability relative to global indexes, since EU Member States share a common legal and cultural framework in which gender equality is widely established as a fundamental right and a development imperative, while the remaining challenges concern persistent unequal opportunities  between women and men and the more difficult access for women and girls to full participation in decision-making and in all spheres of social life. Third, the GEI has been acknowledged as a reliable measurement tool in an audit conducted by the European Commission’s Joint Research Centre \citep{Papadimitriou2020}.

Since 2013 the European Institute for Gender Equality (EIGE) \citep{EIGE} has annually released the GEI, with the purpose of scoring and comparing EU countries based on their levels of gender equality and their progress in reducing gender gaps. The definition and architecture of the GEI are discussed in detail in \cite{BarbieriETal2017}. In this subsection we provide a concise account of its construction, recalling only the essential elements needed to understand our methodological contribution in the following sections. We follow the steps described in Figure~\ref{fig:flow_composite_indicator_simple}.

\paragraph{Step 1: Conceptual framework.}  
The GEI is based on a conceptual framework composed of six domains: Work, Money, Knowledge, Time, Power, and Health, reflecting the multifaceted and latent nature of gender equality in the EU. Each domain is further detailed into 2 or 3 sub-domains, for a total of 14 sub-domains.

\begin{figure}[ht!]
    \centering
    \includegraphics[width=13.97cm]{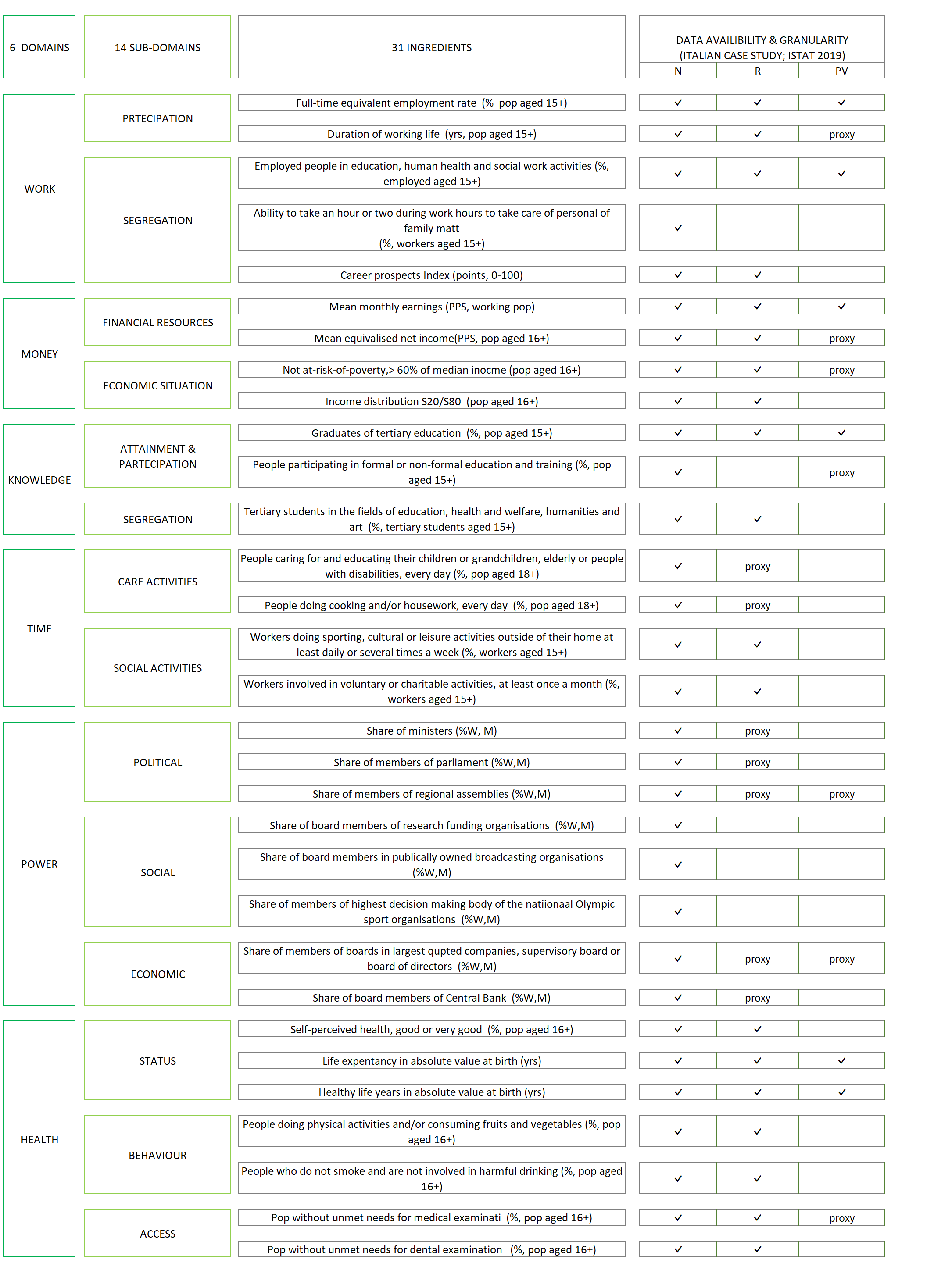}
    \caption{GEI ingredients, organized by subdomains and their corresponding domains. The last column presents a comparison of data availability for Italy (national level, N), Italian regions (sub-national level, R) and provinces (sub-regional level, PV). It should be noted that, at the regional and provincial levels, some ingredients can only be captured through proxy indicators.}
    \label{fig:Fig3}
\end{figure}

\paragraph{Step 2: Operationalization by variables and simple indicators.}  
Within the system of domains and sub-domains, a total of 31 observable variables and simple indicators are selected; henceforth, for brevity and clarity, we refer to them as GEI {\bf ingredients} since they are the main components of the composite indicator. The complete lists of the six domains, fourteen sub-domains, and thirty-one ingredients are shown in the first three columns of Figure~\ref{fig:Fig3}. The fourth column refers to the Italian case study, which will be discussed in Section 4.
Data for ingredients are collected at national level from official and reliable European sources, such as EUROSTAT and EU-SILC \citep{BarbieriETal2017}. Depending on the nature of the variable (categorical, discrete, or continuous), values may be expressed as frequencies, rates, percentages, totals, or averages. For each variable, data are collected separately for women and men. The simple indicator computed at this step is a women-to-average ratio. Formally, for a variable $X_{it}$ (for country $i$ at time $t$), the indicator is given by $x^w_{it}/x^a_{it}$ 
where $x^w_{it}$ is the value for women and $x^a_{it}$ is the average of the female and male values. The gaps between women and men for the variable X for the $i$-th country in the period $t$ period is $ \gamma_{X_{it}}=\left| \frac{x^w_{it}}{x^a_{it}} - 1 \right|$
\paragraph{Step 3: Data processing.}  
The simple indicators produced in the previous step are normalized and re-scaled to calculate a variable-specific gender equality metric ranging from 1 (no gender equality) to 100 (perfect gender equality), defined as:
\begin{equation}
  \Gamma_{X_{it}} = 1 + \left[ \propto_{(X_{it})} \cdot ( 1 - \gamma_{X_{it}}) \right] \cdot 99
  \label{VariableMetric}  
\end{equation}
 where $\propto_{(X_{it})}$ is a country-specific adjusting coefficient to ensure methodological consistency and cross-country comparability \citep{BarbieriETal2017}.

\paragraph{Step 4: Within-domain aggregation.}  
The normalized equality metrics (\ref{VariableMetric}) are first averaged within each sub-domain (nested in domains) using unweighted arithmetic means. Sub-domain values are then aggregated within each domain using an unweighted geometric mean, which reduces the influence of extreme values. This produces a domain-specific GEI score (hereafter domain-GEI).

\paragraph{Step 5: Across-domain aggregation.}  
The final aggregation across all domains is computed as a weighted geometric mean of the six domain-GEIs. Weights, based on expert opinion, are: Work = 0.19, Money = 0.15, Knowledge = 0.22, Time = 0.15, Power = 0.19, and Health = 0.10. The result is rescaled to produce the GEI at the national level on a scale of 0–100, readily interpretable as the percentage of gender equality achieved by a country.

\section{Basics on Bayesian Networks and Object-Oriented Bayesian Networks}

\label{sec3}
In this section Bayesian networks (BNs) are defined, together with their main terminology and properties, providing the necessary foundation for understanding Object-Oriented Bayesian Networks (OOBNs), which extend BNs into a hierarchical structure. Technicalities will be kept to a minimum, focusing solely on the definitions and properties that are fundamental to this article. For more articulated theoretical presentations we refer the reader to \cite{Lauritzen1996}, \cite{cowell1999} and \cite{Koller1997}.

A BN is a  multivariate statistical model that represents sets of conditional dependence and independence relationships,  encoded in a Directed Acyclic Graph (DAG) \citep{Pearl1988, cowell1999}.
A DAG is a pair $G=(V, E)$ where $V$ is the set of nodes and $E$ is the set of directed arcs (arrows) between pairs of nodes. Each node represents a random variable, while missing arrows between nodes imply (conditional) independence between the corresponding variables. A directed graph is acyclic in the sense that it is forbidden to start from a node and, following arrows directions, go back to the starting node. 
Let us consider a DAG for $K$ variables $X_1, X_2, \ldots, X_K$,  $pa(X_j)$ is 
the set of \textit{parents} of $X_j$, i.e. those variables (nodes) with a directed arc pointing to $X_j$. Similarly, $ch(X_j)$ denotes the set of \textit{children} of node $X_j$, i.e. the set of nodes with a directed edge pointing to them from node $X_j$.  

Each node is associated with the conditional probability distribution of the corresponding variable given its parents (if a node has no parents, it is associated with its marginal distribution). A BN is thus a DAG equipped with a joint probability distribution that satisfies the Markov properties \citep{Lauritzen1996}. 
The joint probability distribution can be factorized according to the DAG as the product of the conditional distributions associated to each node given its parents. In general, the chain rule for the distribution of  $(X_1, X_2, \ldots, X_K)$  states that: 
\begin{equation}
 P(X_1, \dots, X_j, \dots, X_K) = \prod_{j} P(X_j \mid pa(X_j))
\label{ChainRuleBN}   
\end{equation}

If the association structure of the phenomenon under consideration is known, the DAG can be built manually on the basis of expert knowledge; if the subject-matter knowledge is partial (i.e. not strong enough to draw the graph), the network structure is necessarily learned from data. The structural learning algorithms can be grouped in three main principal categories: constraint-based, score-based and hybrid algorithms \citep{Kitson2023}.  
Constraint-based algorithms rely on conditional independence tests to learn the dependence structure of the data; score-based algorithms compare networks based on scoring function (e.g., AIC, BIC) and select the structure optimizing this function; hybrid algorithms combine key features of the previous two approaches, relying first on a constraint-based method to reduce the search space and then on a score-based method to select the best structure within that space.

BNs provide an intuitive and transparent explanation of the influence (direct or indirect) of one variable on another. Moreover, BNs are equipped with an inferential engine to make inferences about model parameters, thereby supporting decision makers with an easy-to-read graphical framework and a computationally efficient platform to perform inferential queries and scenario evaluations.

\subsection{From BNs to OOBNs}
In \cite{mecattibayesian2022}, the potential of BNs was explored to improve the monitoring and assessment of gender equality as measured by the GEI. Although the results were promising, the use of a single BN appears somewhat limited with respect to the ambitious goal of interpreting the complex and multilayered nature of both the notion of gender equality and the composite indicator that quantifies it.
To overcome this limitation, we propose the use of OOBNs \citep{Koller1997} as a fundamental tool to both mirror the structure of domains, sub-domains, and variables and model their relationships. This approach takes advantage of a crucial property of BN models known as \textsl{modularity}, which enables the decomposition of a complex multivariate relational structure into simpler sub-networks, while retaining the ability to capture conditional (in)dependencies. In the literature, there are several applications of OOBNs to different fields, such as forensic statistics \citep{dawid2007}, economic decisions \citep{mortera2013} and quality management \citep{musella2015}.

OOBNs are an extension of BNs by which complex and high dimensional problems
can be modelled in terms of inter-related objects. They provide a hierarchical definition of a BN by means of building blocks (instances of classes of BNs). OOBNs are still multivariate statistical models of dependencies among variables, but they allow for more complex modelling than single BNs and require a more articulated system of nodes. In particular, an OOBN is composed of nodes of different nature and scope, the essential being: 
1) {\em chance nodes}, i.e., nodes that represent random variables; 
2) {\em function nodes}, i.e., nodes that are function of their parents;
3) {\em instance nodes}, i.e., objects of the OOBN that represent BNs (or even classes of BNs) and form the hierarchical layers, namely the lower-level(s) of the OOBN; and  
4) {\em interface nodes}, i.e., a special type of (chance or function) nodes that connect the hierarchical levels of the OOBN, channeling information into instances ({\em input} node(s), possibly multiple), and out of instances ({\em output} node, always unique).

\begin{figure}[htp]
\includegraphics[width=15cm]{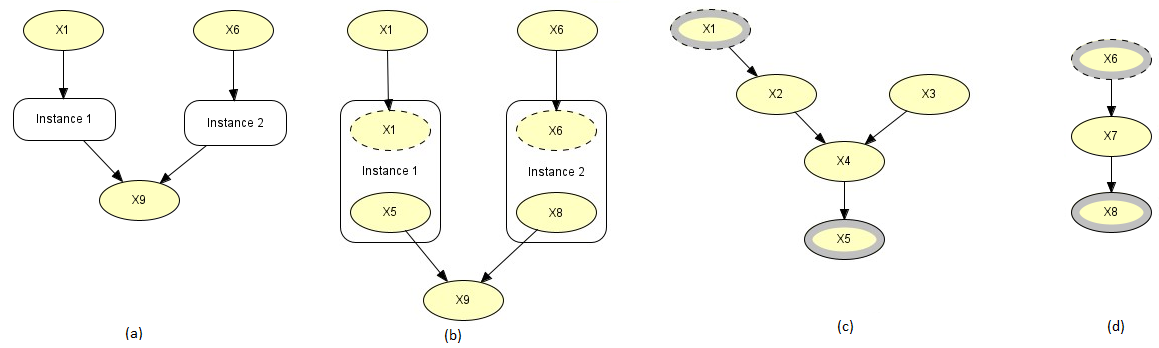}
\caption{Example of OOBN: (a) top-level hyper-network; (b) flow of information from the top-level hyper-network (OOBN) to lower-level sub-networks (BNs); (c) expanded sub-network of Instance 1; (d) expanded sub-network of Instance 2.}
\label{fig:oobn}       
\end{figure}

From now on nodes are indicated in \texttt{teletype} face and classes of networks in \textbf{bold} face.
Figure \ref{fig:oobn}(a) shows an example of the top level network of an OOBN; it comprises three chance nodes (\texttt{X1}, \texttt{X6}, \texttt{X9}), and two instance nodes (or instances, represented as rounded rectangles) of classes \textbf{Instance 1} and \textbf{Instance 2} which are BNs themselves as shown in Figure \ref{fig:oobn}(c) and \ref{fig:oobn}(d) respectively.
Consider for example \ref{fig:oobn}(c). \texttt{X1} and \texttt{X5} are input and output nodes, graphically characterized by an outer grey ring with a dashed line and a full line respectively. In detail \texttt{X1} imports information from the top level network (Figure \ref{fig:oobn}(a)) into \textbf{Instance 1} and \texttt{X5} exports information from \textbf{Instance 1} to the top level network (Figure \ref{fig:oobn}(a)).
Figure \ref{fig:oobn}(b) summarises the entire OOBN and reflects its hierarchical nature. In fact, it provides a picture of the top level network with expanded instances thus 
showing both the chance nodes and the interface (input and output) nodes of the two instances. \\
As for arrows, their meaning may differ from that of statistical dependence when at least one endpoint is not an ordinary chance node. In particular, arrows from chance nodes to input nodes are identity links. So in Figure \ref{fig:oobn}(b) the arrow from the chance node \texttt{X1} to the input node \texttt{X1} of \textbf{Instance 1} represents the fact that information about variable \texttt{X1} is transferred identically within \textbf{Instance 1}. Differently, arrows from output nodes may be either standard statistical dependence links pointing to a chance node, or functional links pointing to a function node associated with an expression depending on its parents.

It is important to note that, since OOBNs are composed of a hierarchy of BNs, all their statistical properties, including the chain rule (\ref{ChainRuleBN}), are retained. Moreover, the inference process of OOBNs is particularly efficient in spite of the complex structure since the standard nodes inside an instance are conditionally independent from the rest of the OOBN, given the interface nodes of that instance. 

In the next section, OOBNs will be used to model the GEI with reference to Italy and, in particular, with a focus on the sub-regional level. 

\section{A case study: the gender equality at Italian provincial level}
\label{sec4}
This section illustrates the added value of using an Object-Oriented Bayesian Network in the construction of a composite indicator, complementing the traditional GEI score by explicitly modeling multivariate relationships \cite{mecattibayesian2022}. The OOBN provides a more faithful representation of the GEI’s layered structure, enabling a richer intersectional perspective and advanced scenario analysis. By visualizing interactions among variables and simulating intervention scenarios, the OOBN supports evidence-based decision-making. An Italian case study is presented to demonstrate the practical contribution of this approach.

 In particular, an OOBN model reflecting the GEI architecture is trained on Italian province-level data and then used both to compare GEI and OOBN in terms of their potential for performing gender impact assessment, and to simulate several illustrative scenarios. Historical factors shaping socio-economic conditions across Italian geographical areas are believed to contribute significantly to the varying degrees of gender inequality. Gender disparities show a tendency to cluster geographically at sub-national level since social-environmental "risk factors" tend to cluster spatially, particularly across Italy. This highlights a need for improved statistical tools to give insights to inform biases' control strategies, to develop locally specific interventions and to prioritize resources allocation across national area.

 Sub-national gender data have been used in literature to compute regional versions of the country-level GEI. For example, \cite{fachelli2023comparative} propose a regional version of the GEI for Italy, France, and Spain, highlighting how different geographical areas possess their own characteristics and require tailored local gender policies. Similarly, \cite{di2021proposing} compute the GEI for Italian regions, underlining the importance of sub-national gender equality measurements to inform policies and prioritize efforts.
Unlike composite indicators, which require a single data point for each dimension covered by the index, learning an OOBN by a data-driven algorithm requires multiple observations for each index dimension. The choice of using micro-regional data is therefore relevant for the model’s implications: finer geographical granularity increases the sample size and generally yields a more reliable model. In Italy, provinces serve as intermediate administrative units between municipalities and regions, with statutory responsibilities in territorial planning, environmental protection, transport, education, and the promotion of equal opportunities. These competences make them key actors in addressing regional heterogeneity and guiding targeted, evidence-based policy implementation.
The remainder of the section is structured as follows. The data used in the model are presented in Section \ref{subsec41}; the statistical model is described in Section \ref{subsec42}; and an illustration of its practical use is provided in Section \ref{subsec43}.

\subsection{Data} 
\label{subsec41}

\begin{figure}[htp]
\includegraphics[width=12cm]{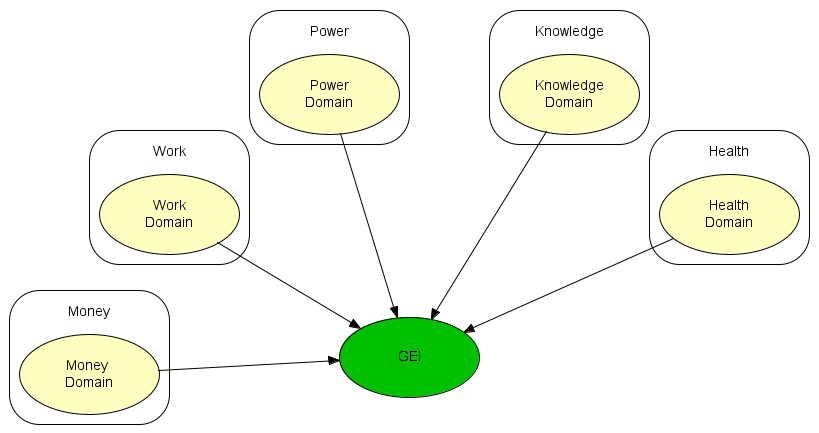}
\caption{Layout of OOBN for GEI - Italian case study}
\label{Fig.6}
\end{figure}

We employ Italian province-level gender data to train an OOBN model for the Italian GEI, referring to the most recent year available, i.e., 2019.
Figure \ref{fig:Fig3} displays a synthesis of GEI ingredient variables, sub-domains, and domains, and compares the availability of data at the national, sub-national, and sub-regional levels. It is important to note that the finer the granularity, the poorer the data availability, which has often required the use of proxy variables. Moreover, data for the Time domain are entirely missing at the provincial level.
However, the modular structure of the OOBN allows for the construction of composite indicators using only the available data, with the flexibility to increase complexity as more data become available. Figure \ref{Fig.6} shows the expected architecture of the GEI calculated on available Italian province-level data as modeled through OOBNs.
Each instance comprises the BN that models the corresponding domain. According to the methodology outlined in Section \ref{subsec23}, the gender gaps $\gamma_{X_{it}}$ for each province $(i = 1, ..., 107)$ at time $t = 2019$ were computed. Once the correction coefficient was applied, these values were transformed into the adjusted gender-gap metrics $\Gamma_{X_{it}}$.

For example, let $X_{it}$ be the ingredient Full-time equivalent employment rate of the Work domain. At time $t=2019$ and for the province $i = \text{Turin}$, the indicator takes the following values: $\Tilde{x}^m_{\text{Turin},2019} = 71.23$, $\Tilde{x}^w_{\text{Turin},2019} = 59.62$, and $\Tilde{x}^a_{\text{Turin},2019} = 65.43$ (source: ISTAT, Labour Force Survey). The gender gap is measured by:$$\gamma_{(X_{it})} = \left| \frac{59.62}{65.43}-1\right| = 0.09.$$

The correcting factor $\propto_{(X_{it})}$ equals 0.94 for the province of Turin in 2019. Consequently, the gender-gap metric is:
$$\Gamma_{(X_{Turin,2019})}=1+[0.94\cdot(1-0.09)]\cdot99 = 85.76$$
indicating a relatively high degree of proximity to gender equality in employment for this province.

In order to better represent the geo-socio-economic context in which gender balancing policies are to be conceived and actions taken, the OOBN has been embedded within the broader national framework through the identification and inclusion of additional relevant variables, henceforth {\it extra variables}. 
These variables capture key aspects of the socioeconomic structure of the area and are easily accommodated within the OOBN model to build a more inclusive picture showing how gender-related indicators are inherently linked to social and economic features. The rich insight offered by this extended model is essential for designing and implementing effective gender-equality policies. A thorough review of the gender-gap literature informed the selection of additional variables not directly included in the GEI but strongly connected to its core domains. For instance \cite{bettioGenderSegregationLabour2009, blauGenderWageGap2017, gauchatOccupationalGenderSegregation2012} highlight key drivers of gender segregation such as the dominant economic sector, education levels, caregiving responsibilities, legal barriers, and cultural stereotypes. \cite{azmatGenderGapsUnemployment2006, chevalierEducationOccupationCareer2007} also emphasize the impact of fertility, childcare, and legal constraints on women’s employment, as well as the role of discrimination and personal choice in educational and occupational outcomes. Health disparities related to caregiving and geography, along with knowledge-related factors such as parental education and female teacher representation, further contribute to the gender gap \citep{riphahnWhatDrivesReversal2015, eekGenderInequalityHome2015}. For the dimension of power, cultural background, family context, and systemic discrimination are identified as critical influences \citep{lombardoGenderInequalityPolitics2008, foxUncoveringOriginsGender2014}. Some province-level controls such as GDP, geographical area, and population density were included as significant characterizations of Italian provinces.
However, the set of relevant extra-variables remains limited due to the unavailability of data at the provincial level.

Tables \ref{tab:GEI_ingredients} and \ref{tab:extra_variables} summarize all the data included in the model, describing respectively ingredient variables (ordered according to their domain, see Figure \ref{fig:Fig3}) and extra variables (listed alphabetically). Specifically, in both tables, the first column lists the data labels that also represent node names in the OOBN model; the second column provides the data description; the third column reports the data range; and the last column indicates the data source.

\begin{table}
\centering
\footnotesize
\renewcommand{\arraystretch}{1.15}
\begin{tabularx}{\textwidth}{l X r X}
\textbf{Data/Node name} & \textbf{Description} & \textbf{Range} & \textbf{Survey \& Source} \\
\hline
Employment rate equality &
Full-time equivalent employment rate (percentage of population 15+). &
1--100 &
Labour Force Survey by ISTAT. \\

Active rate equality &
Number of people expected to be active in the labour market; proxy of duration of working time. &
1--100 &
Labour Force Survey by ISTAT. \\

Work segregation equality &
Employed people in education, human health, and social work activities (\%, 15+ employed). &
1--100 &
Well-being and sustainability indicators by ISTAT. \\

Income equality &
Mean monthly earnings (PPS, working population). &
1--100 &
Well-being and sustainability indicators by ISTAT. \\

Saving equality &
Average per capita amount of retirement income; proxy of mean equivalent net income (PPS, 16+ population). &
1--100 &
Well-being and sustainability indicators by ISTAT. \\

Non-poverty equality &
Percentage of retirees with medium/high retirement income; proxy of not being at risk of poverty, i.e., $\geq$ 60\% of median income (\%, 16+ population). &
1--100 &
Well-being and sustainability indicators by ISTAT. \\

Education equality &
Graduates and tertiary education (\%, 15+ population). &
1--100 &
Ministry for Education, University and Research. \\

Education participation equality &
Percentage of new graduates enrolling for the first time at university in the same year in which they obtained their high school degree; proxy of participation in education and training. &
1--100 &
Well-being and sustainability indicators by ISTAT. \\

Political power equality &
Percentage of women in local administrations; proxy for share of members of regional assemblies (\% women, men). &
1--100 &
Ministry of the Interior. \\

Economic power equality &
Number of women enterprises in the province; proxy for share of members of boards in largest quoted companies, supervisory board, or board of directors (\% women, men). &
1--100 &
Chamber of commerce. \\

Health status equality &
Life expectancy at birth (average number of years a person aged 65+ is expected to live without any limitation). &
1--100 &
Well-being and sustainability indicators by ISTAT. \\

Health access equality &
Avoidable mortality; proxy for population without unmet needs for medical examination. &
1--100 &
Well-being and sustainability indicators by ISTAT. \\
\hline
\end{tabularx}
\caption{GEI ingredients included in the model.}
\label{tab:GEI_ingredients}
\end{table}

\begin{table}
\centering
\footnotesize
\renewcommand{\arraystretch}{1.15}
\begin{tabularx}{\textwidth}{l X r X}
\textbf{Data/Node name} & \textbf{Description} & \textbf{Range} & \textbf{Survey \& Source} \\
\hline
Average number of children &
Average number of children per woman. &
0.92--1.71 &
Fertility indicators by ISTAT. \\

Density &
Population density in the province. &
36.73--2585.55 &
Population Census by ISTAT. \\

Employment rate &
Full-time equivalent employment rate (percentage of population 15+). &
35.84\%--74.05\% &
Labour Force Survey by ISTAT. \\

Firm size &
Average number of persons employed in active enterprises. &
2.27--6.7 &
Statistical Register of Active Enterprises (ASIA). \\

GDP &
GDP at province level. &
14056.53--49884.09 &
National accounts main aggregates by ISTAT. \\

Geographical area &
Territory macro-area of the province (NO = Nord Ovest; NE = Nord Est; C = Center; S = South; I = Islands). &
-- &
Administrative unit boundaries. \\

Internet access &
Percentage of residents with broadband internet access. &
4.2\%--26.59\% &
Territorial development indicators by ISTAT. \\

Higher education rate &
Proportion of graduates among residents. &
0.002--0.008 &
Ministry for Education, University and Research. \\
\hline
\end{tabularx}
\caption{Extra variables included in the model.}
\label{tab:extra_variables}
\end{table}

The OOBN model was fitted using the \texttt{Hugin Expert} software. All variables were preliminarily categorized into empirical quartiles. Their relationships were then estimated using the NPC algorithm \citep{steckBayesianBeliefNetworks1999}, an extension of the PC algorithm that  effectively tailors the learning process, and thus the resulting estimated OOBN, to the specific requirements of our case study. The NPC algorithm is particularly well-suited for limited data sets and facilitates recursive testing of marginal and conditional associations between categorical variables while incorporating logical constraints. During the learning phase, users can guide the process by prohibiting or specifying certain edge directions, enforcing hierarchical relationships among nodes, and managing interdependent uncertain links, where the presence or direction of one arc depends on another. Such ambiguities can be resolved by integrating subject-matter expertise. In essence, the algorithm allows users to choose, among independence-equivalent models, the one that best aligns with the problem under study by combining automated learning with expert knowledge. Once the conditional independence structure among variables has been established, parameter estimation proceeds by estimating the conditional probability tables associated with the network’s nodes.

\subsection{An OOBN Model for GEI at the Italian Province level} 
\label{subsec42} 
\begin{figure}[htp]
\includegraphics[width=15cm]{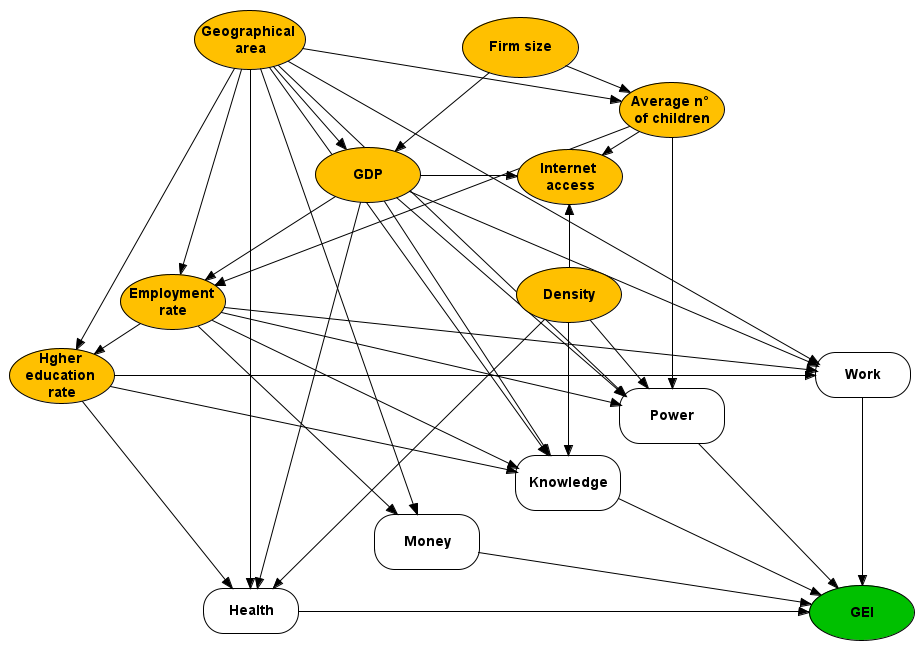}
\caption{Top-level network of the OOBN model for GEI at the Italian province level}
\label{fig:Fig12}      
\end{figure}

Figure \ref{fig:Fig12} shows the learned top level of the OOBN for the GEI at the Italian provincial level.
Model estimation followed a multi-stage procedure. In the first stage, an exploratory analysis was carried out: a fully data-driven model, initially free from any predefined structure, was estimated to explore associations among extra variables and ingredients. Specifically, the NPC algorithm was employed to investigate these relationships while allowing for certain structural constraints during the learning process. For example, it was assumed that the extra variables held a preliminary position with respect to the gender perspective. Consequently, during the estimation, they were constrained so as not to be influenced by the GEI ingredient variables. This stage enabled the data-driven identification of those extra variables serving as input variables for BN instances. The selection was based on the observed interrelations between extra variables and GEI ingredients that emerged during this exploratory phase, prior to the formal establishment of the GEI architecture via OOBN. 

In the second stage, each BN instance (one per domain) was learned from data, including both the variables inherently associated with the domain and the extra variables identified as inputs in the previous step (see Section \ref{subsec:Instances} for details). The model architecture was then finalized by introducing function nodes both for the domains and subdomains within the BN instances and for the overall GEI Indicator.

In particular, all BN instances ultimately converge, through the output nodes, to {\tt GEI}, the OOBN model’s target node. 
{\tt GEI} is a function node whose values are computed using a weighted geometric mean. To compensate for the unavailable Time domain, the original weights were normalized as follows: 0.22 for Work, 0.18 for Money, 0.26 for Knowledge, 0.12 for Power, and 0.22 for Health. The formula is
$$
GEI=\tt{Work\ Domain}^{0.22}*\tt{Money\ Domain}^{0.18}*\tt{Knowledge\ Domain}^{0.26}*\tt{Power\ Domain}^{0.12}*\tt{Health\ Domain}^{0.22}
$$
where \texttt{Work Domain, Money Domain, Knowledge Domain, Power Domain and Health Domain} are the output nodes of the corresponding BN instances serving to propagate the domain scores to the \texttt{GEI} node at the top level.

Generally speaking, the goodness of fit assessment of BNs involves evaluating how accurately the probabilistic counterpart of the OOBN represents the underlying data distribution from which it was derived. This step is crucial to validate the model’s reliability and its predictive accuracy, ensuring that the inferred dependencies and conditional probability tables accurately capture the true relationships among variables. In line with this, the overall goodness-of-fit of the model was assessed by generating a sample of 500 data points from the learned network and subsequently evaluating how well these data conform to the learned network structure. 

In detail, the True Positive Rate (TPR) and False Positive Rate (FPR) are employed to evaluate the classification performance of the BN. The TPR measures the model’s ability to correctly identify positive instances, while the FPR quantifies the tendency of the model to incorrectly classify negative instances as positive. These metrics offer valuable insight into the trade-off between sensitivity and specificity, and are used to assess the network’s effectiveness in accurately classifying cases and, consequently, in predicting GEI levels. Predictive performance was then evaluated using the  Receiver Operating Characteristic (ROC) curve \citep{metz1978roc}, which plots TPR against FPR. The Area Under the Curve (AUC) summarizes the discriminant power, ranging from 0.5 (no discrimination) to 1 (perfect discrimination). To predict the highest level of GEI, the model achieved an AUC score greater than 0.90, which, according to the classification by \citep{swets1988accuracy}, indicates optimal predictive accuracy. 

In the following Section, a detailed presentation of the BN instances is provided.

\subsubsection{Modeling domains by BN instances}
\label{subsec:Instances}
\begin{figure}[htbp]
    \centering

    \begin{minipage}[t]{0.45\textwidth}
        \centering
        \includegraphics[width=\linewidth]{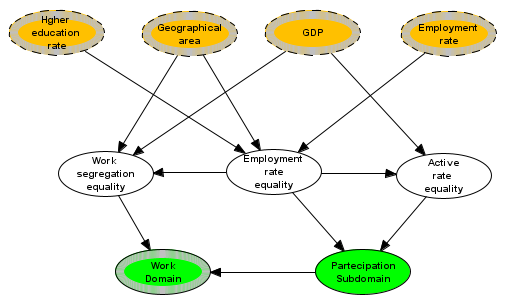}
        \par\smallskip
        \textbf{(a)}
    \end{minipage}
    \hfill
    \begin{minipage}[t]{0.40\textwidth}
        \centering
        \includegraphics[width=\linewidth]{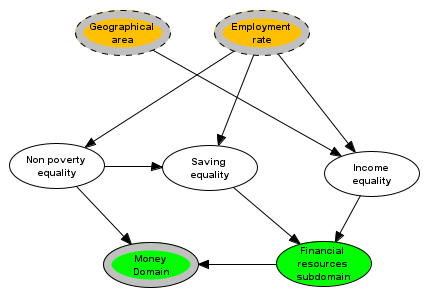}
        \par\smallskip
        \textbf{(b)}
    \end{minipage}

    \par\medskip

    \begin{minipage}[t]{0.45\textwidth}
        \centering
        \includegraphics[width=\linewidth]{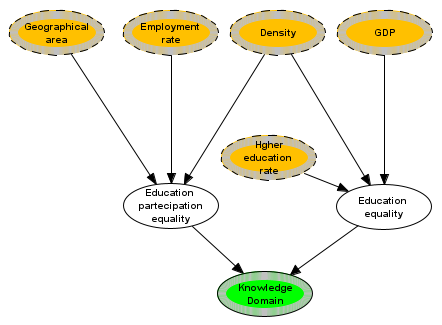}
        \par\smallskip
        \textbf{(c)}
    \end{minipage}
    \hfill
    \begin{minipage}[t]{0.45\textwidth}
        \centering
        \includegraphics[width=\linewidth]{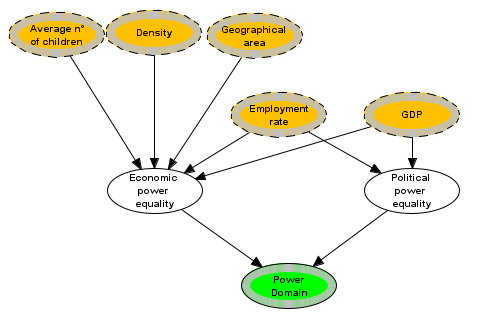}
        \par\smallskip
        \textbf{(d)}
    \end{minipage}

    \par\medskip

    \begin{minipage}[t]{0.50\textwidth}
        \centering
        \includegraphics[width=\linewidth]{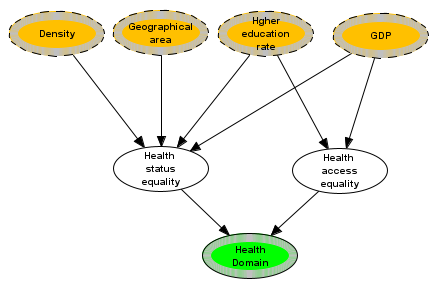}
        \par\smallskip
        \textbf{(e)}
    \end{minipage}

    \caption{BN instances for the domains:
    (a) BN for the \textbf{Work} domain;
    (b) BN for the \textbf{Money} domain;
    (c) BN for the \textbf{Knowledge} domain;
    (d) BN for the \textbf{Power} domain; and
    (e) BN for the \textbf{Health} domain.}
    \label{fig:CombinedCaption}
\end{figure}

In this Section, the BN instances learned for each domain are presented. Each BN instance represents a single GEI domain and consists of input nodes coloured in orange, which always represent extra variables (see Section \ref{subsec42}); GEI ingredients shown as white chance nodes, and a green output node representing the domain score, calculated using the official methodology. This involves taking an unweighted geometric mean of the sub-domain scores (when applicable), which are in turn the unweighted arithmetic mean of the ingredients involved. Figure~\ref{fig:CombinedCaption} consists of five panels, each showing the estimated BN instance for a GEI domain, based on data available at the Italian provincial level.

In particular, Figure~\ref{fig:CombinedCaption}(a) shows the BN for the \textbf{Work} domain. In this domain, the extra input variables are {\tt Higher education rate}, {\tt Geographical area}, {\tt GDP}, and {\tt Employment rate}, indicating that the domain  is influenced by these characteristics that describe the Italian socio-economic system. Both the \texttt{Participation Subdomain} and the \texttt{Work Domain} are deterministic nodes (green), whose values are calculated using expressions involving their respective parents.\\
The BN for the \textbf{Money} domain is shown in Figure~\ref{fig:CombinedCaption}(b). The input variables for this domain include \texttt{Geographical area} and the \texttt{Employment rate} at the provincial level. This aligns with published  literature, which highlights regional disparities in the Italian labor market and their influence on economic well-being (see, for instance, \cite{tonutti2024inwork}). The \texttt{Employment rate} plays a central role, being directly connected to all core components of the domain, while the \texttt{Geographical area} is directly linked to \texttt{Income equality}. Both the \texttt{Financial resources Subdomain} and the \texttt{Money Domain} are deterministic nodes, with their values computed through expressions involving their parents, in accordance with GEI guidelines.\\
Figure~\ref{fig:CombinedCaption}(c) presents the BN for the \textbf{Knowledge} domain. As in the previous cases, several input variables (resulting from the structural learning process) are included, capturing ungendered higher education levels — {\tt Higher education rate} — geographical characteristics — such as \texttt{Geographical area} and \texttt{Density} — and economic factors — such as \texttt{GDP} and \texttt{Employment rate}. Notably, the overall state of the labour market, reflected in the employment rate, influences educational participation from a gender perspective, revealing patterns of gender segregation. The \texttt{Knowledge Domain} node represents the domain score, calculated according to the GEI methodology.\\
Figure~\ref{fig:CombinedCaption}(d) shows the BN for the \textbf{Power} domain. As in the previous cases, the input nodes in this instance include \texttt{Geographical area}, \texttt{Density}, \texttt{GDP}, and \texttt{Employment rate}. Another input variable in this domain is the \texttt{Average number of children}, capturing the dual role women often assume in both care and economic spheres. Structural learning reveals that \texttt{Political power} and \texttt{Economic power} are conditionally independent given the \texttt{Employment rate}, highlighting the influence of labour market participation on access to socio-political power. The \texttt{Power Domain} score is calculated based on its constituent variables, \texttt{Political power} and \texttt{Economic power}, in accordance with GEI guidelines.\\
Finally, Figure~\ref{fig:CombinedCaption}(e) illustrates the BN for the \textbf{Health} domain. The extra variables \texttt{Density}, \texttt{Geographical area}, \texttt{Higher education rate}, and \texttt{GDP} serve as input nodes in the model. These variables reflect both the geographical disparities across the Italian provinces and the influence of economic resources on access to healthcare. Additionally, \texttt{Health status equality} and \texttt{Health access equality} are conditionally independent given the \texttt{Higher education rate}, indicating a connection between educational attainment and equitable health outcomes. The \texttt{Health Domain} node is associated with a table whose values are given by an expression integrating \texttt{Health status equality} and \texttt{Health access equality}, in accordance with GEI guidelines.\\
In the next Section, the usage of the OOBN is shown by simulating different scenarios. 

\subsection{A tool for what-if analysis}  
\label{subsec43}

By its inferential engine, the OOBN can serve for assessing potential policies aimed at enhancing the GEI. The model in fact allows to  simulate an action on one or more variables, belonging to the same or different BN instances, and investigate the effect of this action on the remaining variables of the model. Information instantiated in one or more nodes is propagated through the network, updating all probability distributions.

\begin{figure}[htp]
\centering
\includegraphics[width=12cm]{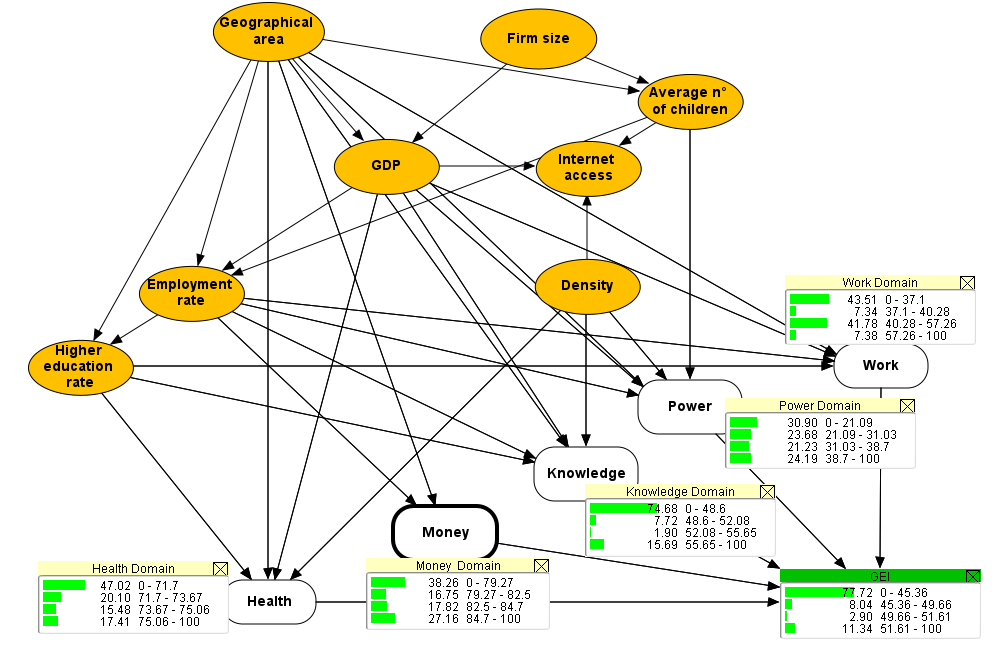}
\caption{OOBN model for GEI with marginal probability distributions of GEI and domain nodes}
\label{fig:OL_baseline} 
\end{figure}

Firstly, a baseline scenario where none of the nodes are instantiated is considered. The scenario, shown in Figure \ref{fig:OL_baseline}, displays the probability distributions of the GEI and of its five domains.

The observed distributions reflect the information contained in Italian province-level data and highlight how Italy is still far from achieving gender parity, with a probability of 0.78 associated to the lowest GEI level and a probability of 0.11 associated to its highest level.

\begin{figure}[htp]
\centering
\includegraphics[width=12cm]{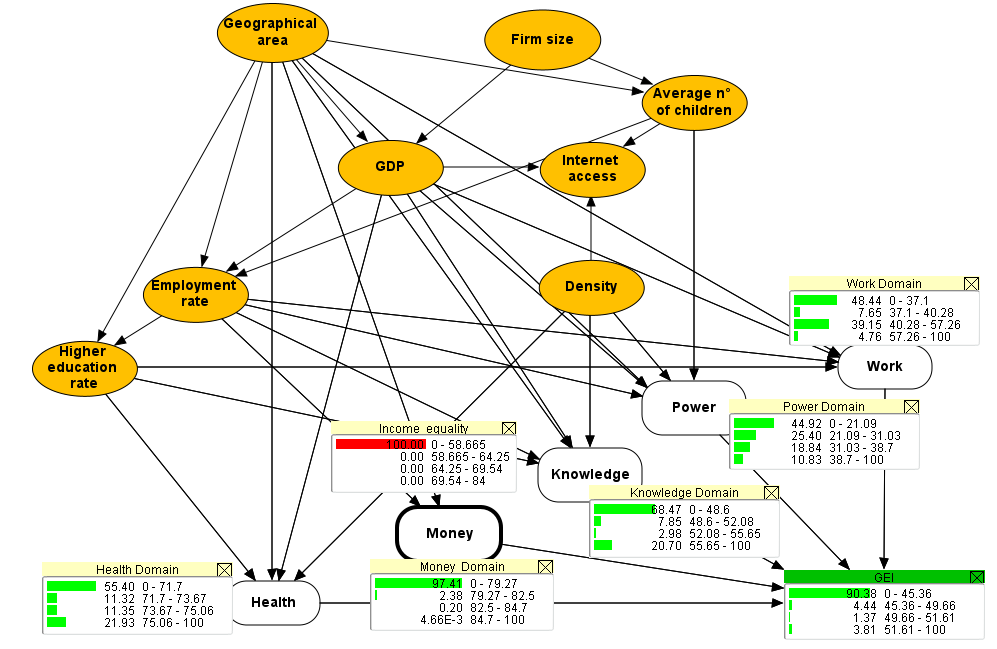}
\caption{OOBN model for GEI marginal probability distributions of GEI and domain nodes with low income equality scenario instantiated}
\label{fig:OL_inc_low} 
\end{figure}

We suppose to observe a very low income equality level, an ingredient belonging to the Money domain. This new evidence is then inserted into the baseline scenario and changes to the displayed probability distributions are investigated. Therefore, the node \texttt{Income equality} is instantiated to its lowest level and the OOBN probability distributions are updated accordingly.
Simulating a scenario where income equality is very low reflects the idea of investigating how GEI domains, ingredients and socio-economic extra-variables behave when provinces face extreme income inequality conditions. 
Figure \ref{fig:OL_inc_low} shows the obtained results.

Consistently with the GEI architecture, the probability distribution of \texttt{Money Domain} shows a very high probability (0.97) associated to its lowest state, reflecting how increasing the inequality captured by an ingredient cascades on the whole domain. The substantial shift of this probability distribution is consistent with the idea that when one ingredient worsens, other ingredients of the same domain follow, reinforcing the negative effect on the domain distribution. Furthermore, almost all the other domain distributions shift toward their lowest levels, indicating that lower income equality is generally linked to worse conditions across the GEI dimensions. This reflects the fact that that areas with very low income equality often lack  the socio-economic conditions required to sustain broader equal opportunities. When we investigate the distribution of the \texttt{GEI} node resulting from this scenario, we observe a probability of 0.91 and 0.04, linked respectively to its lowest and highest levels. Remarkably, these are the result of a systemic effect and not only of the direct effect that changing income equality has on the GEI computation. Low income equality in fact, often arises in situations where women face greater economic vulnerability, and this can easily extend  to other dimensions, such as career progression, access to services, or participation in public life, helping to explain the system-wide deterioration captured by the OOBN in this scenario.

\begin{figure}[htp]
\centering
\includegraphics[width=12cm]{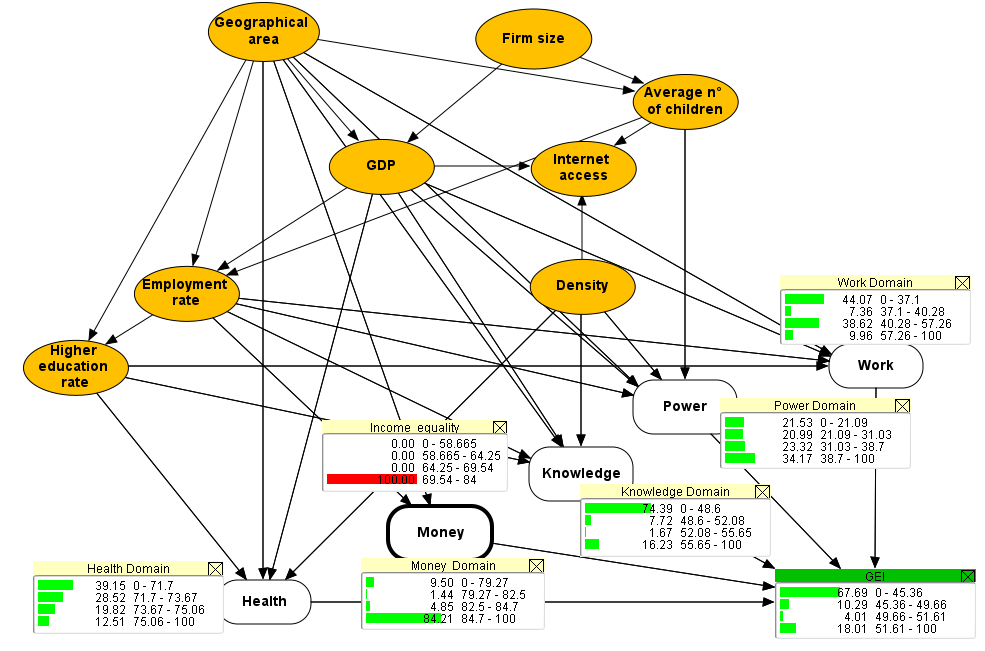}
\caption{OOBN model for GEI marginal probability distributions of GEI and domain nodes with high income equality scenario instantiated}
\label{fig:OL_inc_high}
\end{figure}

Figure \ref{fig:OL_inc_high} shows instead the scenario where \texttt{income equality} is instantiated to its highest level. Simulating this scenario, corresponds to investigating what happens to the system when the income gender-gap is very low.

Analogously to what has been observed in the previous scenario, when evidence on a node is provided, information propagates across the whole network affecting \texttt{GEI} not only directly, but also through all the other variables in the model, fully exploiting the multivariate vocation of OOBNs. Setting \texttt{income equality} to its highest level is in fact associated with improved conditions in nearly all other domains, indicating that enhancements in one dimension of gender equality tend to translate into positive effects on the others. This aligns with the idea that when economic conditions become more equal, women often gain better access to opportunities related to work, knowledge, power, and health, producing improvements that reinforce one another rather than remaining isolated.
Consequently, the GEI distribution substantially shifts towards its positive levels, with respect to the previous scenario. In particular, the probability associated to the lowest level of the \texttt{GEI} decreases from 0.91 to 0.68 denoting a 25\% variation. On the other hand, the probability of observing the highest level of \texttt{GEI} shows a substantial increase, shifting from 0.04 to 0.18.
These changes suggest that improving income equality can act as a catalyst, lifting the overall gender-equality profile of a territory, and highlighting how interconnected the different dimensions of the GEI are.

In order to facilitate a comparison between the OOBN model and the traditional GEI score, and to highlight the added value offered by the former over  the latter, the same simulations are then conducted on a \textit{constrained} version of the OOBN, imitating the functioning of a composite indicator.
Specifically, the constrained OOBN is obtained by imposing that each ingredient takes a certain, fixed level, actually blocking all the paths both between ingredients and between ingredients and extra-variables. Consequently, the information can be propagated only from each ingredient to its reference domain and in turn to the GEI score, following the
structure of the traditional indicator represented in Figure \ref{Fig.6}. 

Once each ingredient node has been istantiated to a certain level, the network and obtained probability distributions are then employed as a new baseline to re-assess the income equality scenario, keeping every other ingredient level fixed. Here, we instantiated each ingredient to its second quartile with the aim of achieving simplicity and staying consistent with the learned ingredient probability distributions.
As already mentioned, this use of the OOBN mimics a traditional composite indicator approach where simulating the variation of a single ingredient would directly result in a proportional change in the GEI level.

Table \ref{tab:Tab2} reports the probability distributions of the \texttt{GEI} and its domains obtained from the scenarios applied to the \textit{unconstrained} OOBN (columns 3-5), and to the constrained OOBN (columns 6-8).

\begin{table}[htbp]
\centering
\small
\setlength{\tabcolsep}{3.5pt}
\renewcommand{\arraystretch}{1.1}

\begin{tabular}{ll rrr rrr}
\hline
\textbf{Node name} & \textbf{Level} &
\textbf{\makecell{Unconstrained\\OOBN\\baseline}} &
\textbf{\makecell{Unconstrained\\OOBN\\low income}} &
\textbf{\makecell{Unconstrained\\OOBN\\high income}} &
\textbf{\makecell{Constrained\\OOBN\\baseline}} &
\textbf{\makecell{Constrained\\OOBN\\low income}} &
\textbf{\makecell{Constrained\\OOBN\\high income}} \\
\hline
Work domain & 0--37.1 & 43.51 & 48.44 & 44.07 & 46.00 & 46.00 & 46.00 \\
Work domain & 37.1--40.28 & 7.34 & 7.65 & 7.36 & 25.57 & 25.57 & 25.57 \\
Work domain & 40.28--57.26 & 41.78 & 39.15 & 38.62 & 28.43 & 28.43 & 28.43 \\
Work domain & 57.26--100 & 7.38 & 4.76 & 9.96 & 0.00 & 0.00 & 0.00 \\[2pt]

Money domain & 0--79.27 & 38.26 & 97.41 & 9.50 & 0.00 & 98.09 & 0.00 \\
Money domain & 79.27--82.5 & 16.75 & 2.38 & 1.44 & 100.00 & 1.91 & 3.26 \\
Money domain & 82.5--84.7 & 17.82 & 0.20 & 4.85 & 0.00 & 0.00 & 14.85 \\
Money domain & 84.7--100 & 27.16 & 0.00 & 84.21 & 0.00 & 0.00 & 81.90 \\[2pt]

Knowledge domain & 0--48.6 & 74.68 & 68.47 & 74.39 & 100.00 & 100.00 & 100.00 \\
Knowledge domain & 48.6--52.08 & 7.72 & 7.85 & 7.72 & 0.00 & 0.00 & 0.00 \\
Knowledge domain & 52.08--55.65 & 1.90 & 2.98 & 1.67 & 0.00 & 0.00 & 0.00 \\
Knowledge domain & 55.65--100 & 15.69 & 20.70 & 16.23 & 0.00 & 0.00 & 0.00 \\[2pt]

Power domain & 0--21.09 & 30.90 & 44.92 & 21.53 & 7.84 & 7.84 & 7.84 \\
Power domain & 21.09--31.03 & 23.68 & 25.40 & 20.99 & 84.00 & 84.00 & 84.00 \\
Power domain & 31.03--38.7 & 21.23 & 18.84 & 23.32 & 8.16 & 8.16 & 8.16 \\
Power domain & 38.7--100 & 24.19 & 10.83 & 34.17 & 0.00 & 0.00 & 0.00 \\[2pt]

Health domain & 0--71.7 & 47.02 & 55.40 & 39.15 & 3.36 & 3.36 & 3.36 \\
Health domain & 71.7--73.67 & 20.10 & 11.32 & 28.52 & 96.64 & 96.64 & 96.64 \\
Health domain & 73.67--75.06 & 15.48 & 11.35 & 19.82 & 0.00 & 0.00 & 0.00 \\
Health domain & 75.06--100 & 17.41 & 21.93 & 12.51 & 0.00 & 0.00 & 0.00 \\[2pt]

GEI & 0--45.36 & 77.72 & 90.83 & 67.69 & 91.06 & 98.24 & 87.89 \\
GEI & 45.36--49.66 & 8.04 & 4.44 & 10.29 & 8.10 & 1.67 & 10.33 \\
GEI & 49.66--51.61 & 2.90 & 1.37 & 4.01 & 0.71 & 0.08 & 1.37 \\
GEI & 51.61--100 & 11.34 & 3.81 & 18.01 & 0.13 & 0.01 & 0.42 \\
\hline
\end{tabular}

\caption{Probability distributions (\%) across different OOBN scenarios.}
\label{tab:Tab2}
\end{table}

We start by considering the probability distributions of \texttt{GEI} and its domains when \texttt{income equality} is set to its lowest level, keeping all the other ingredient levels fixed, shown in Table \ref{tab:Tab2}.

Instantiating \texttt{income equality} to its lowest state in the constrained OOBN directly shifts the probability distribution of \texttt{Money Domain} towards its lowest level. However, since the paths between ingredients and extra-variables are blocked in the constrained network, the other domains are not affected, contrarily to what we observed in the unconstrained OOBN simulation. In other words, when the OOBN is used as if it were a traditional composite indicator, indirect interaction channels are shut down and information does not flow across domains. Consequently, changes in the distribution of an ingredient are assessable only within the domain to which that ingredient belongs. Following the architecture of the composite indicator, the distributional shift of the \texttt{Money Domain} affects the GEI node in turn, increasing the probability associated with its lowest level from 0.91 to 0.98. Conversely, the probabilities linked to its highest levels decrease, consistently with the composite indicator’s computation. Unlike in the unconstrained OOBN scenario, the change in the GEI distribution resulting from instantiating \texttt{income equality} to its lowest level in the constrained OOBN does not capture the systemic effects produced by indirect interactions. Instead, it merely reflects the changes implied by the GEI’s computational structure. In practice, this means that the constrained OOBN cannot reflect how difficulties in one area often reinforce difficulties in others, an aspect that is essential for understanding real territorial patterns of gender inequality.

Subsequently, we consider the scenario in which \texttt{income equality} is set to its highest level. The results are shown in the relative column of Table \ref{tab:Tab2}.
Consistently with the model architecture, the probability distribution of the \texttt{Money domain} shifts towards its right tail, with the probability of its highest level increasing to 0.82 and the probability linked to its lowest level decreasing to 0. In turn, the probability associated to the lowest level of the \texttt{GEI} decreases by 10 percentage points, going from 0.98 to 0.88, while its highest level probability shows a slight increase, from almost 0 to 0.004.
As noted above, these results deterministically derive from the computational structure of the GEI score. 

Comparing the variation in the \texttt{GEI} distribution obtained employing the unconstrained and constrained OOBNs reveals that the latter may only offer a partial view of how the indicator responds to changes in its ingredients. The unconstrained OOBN model on the other hand, highlights how the complex relational structure connecting extra-variables, ingredients, and domains is essential for accurately measuring and promoting gender equality by a fully multivariate/intersectional analysis.

Lastly, a scenario that involves both extra variables and GEI ingredients is evaluated on the baseline OOBN  of Figure \ref{fig:OL_baseline}.
This evaluation leverages the unconstrained OOBN’s capacity to embed the GEI computation within the Italian socio-economic context. In this framework, scenario simulations can extend beyond changes in the GEI ingredients to include variations in other social and economic characteristics of the territories.

Suppose, for example, that policymakers are interested in how a policy that simultaneously increases ungendered employment and work segregation equality affects the  gender equality level in the country, as measured by the  GEI. Imposing high segregation equality together with high employment levels aims to ensure that, as employment rises, jobs across different business areas are distributed equally between women and men.
To simulate such a policy, we instantiate \texttt{Work segregation equality} and \texttt{Employment rate} to their maximum level. Moreover, given the high level of geographical heterogeneity that characterizes Italian regions, policy makers may also be interested in how the policy would affect different areas. In order to accomplish this goal, the node \texttt{Geographical area} is instantiated to two exemplificative Italian macro-regions, namely North-East and South, and the results are compared. The obtained distributions of the GEI and its domains are reported in Table \ref{tab:extra_scen}, together with the relative baselines.

\begin{table}[htbp]
\centering
\small
\setlength{\tabcolsep}{4pt}
\renewcommand{\arraystretch}{1.1}

\begin{tabular}{ll rrrr}
\hline
\textbf{Node name} & \textbf{Level} &
\textbf{\makecell{North-East\\baseline}} &
\textbf{\makecell{North-East\\scenario}} &
\textbf{\makecell{South\\baseline}} &
\textbf{\makecell{South\\scenario}} \\
\hline
Work domain & 0--37.1 & 37.78 & 1.46 & 50.19 & 9.75 \\
Work domain & 37.1--40.28 & 7.48 & 0.26 & 7.85 & 1.73 \\
Work domain & 40.28--57.26 & 43.08 & 7.10 & 37.54 & 26.56 \\
Work domain & 57.26--100 & 11.67 & 91.18 & 4.42 & 61.96 \\[2pt]

Money domain & 0--79.27 & 15.23 & 2.93 & 77.38 & 24.52 \\
Money domain & 79.27--82.5 & 8.27 & 3.95 & 15.17 & 14.75 \\
Money domain & 82.5--84.7 & 27.66 & 26.42 & 4.92 & 24.12 \\
Money domain & 84.7--100 & 48.85 & 66.70 & 2.54 & 36.61 \\[2pt]

Knowledge domain & 0--48.6 & 67.73 & 61.24 & 78.35 & 72.19 \\
Knowledge domain & 48.6--52.08 & 8.33 & 11.12 & 6.78 & 7.12 \\
Knowledge domain & 52.08--55.65 & 2.23 & 2.55 & 1.79 & 2.12 \\
Knowledge domain & 55.65--100 & 21.71 & 25.09 & 13.08 & 18.57 \\[2pt]

Power domain & 0--21.09 & 25.93 & 25.96 & 42.23 & 12.01 \\
Power domain & 21.09--31.03 & 22.18 & 22.92 & 26.05 & 17.61 \\
Power domain & 31.03--38.7 & 21.06 & 21.61 & 19.78 & 27.28 \\
Power domain & 38.7--100 & 30.83 & 29.51 & 11.94 & 43.10 \\[2pt]

Health domain & 0--71.7 & 44.25 & 43.76 & 56.19 & 40.26 \\
Health domain & 71.7--73.67 & 23.98 & 22.15 & 8.94 & 23.68 \\
Health domain & 73.67--75.06 & 19.76 & 26.27 & 13.07 & 17.15 \\
Health domain & 75.06--100 & 12.00 & 7.82 & 21.80 & 18.91 \\[2pt]

GEI & 0--50.03 & 55.73 & 41.96 & 86.87 & 49.14 \\
GEI & 50.03--53.05 & 10.15 & 10.69 & 4.38 & 11.17 \\
GEI & 53.05--54.7 & 5.11 & 4.98 & 1.85 & 4.95 \\
GEI & 54.7--100 & 29.01 & 42.37 & 6.90 & 34.79 \\
\hline
\end{tabular}

\caption{Probability distributions (\%) across scenarios comparing North-East and South regions, with corresponding baseline distributions.}
\label{tab:extra_scen}
\end{table}

The scenario applied to the North-Est macro-region shows how increasing the employment rate and work segregation equality, positively affects \texttt{Money Domain} and \texttt{Work Domain}, with the respective distributions shifting towards the highest equality levels. Note that while the distribution of \texttt{Work Domain} directly benefits of setting one of its ingredients, namely \texttt{Work segregation equality}, to its maximum level, \texttt{Money Domain} distribution improves as a result of the structure of direct and indirect associations emerging from the OOBN model. The mentioned variations in the domains, in turn, affect the distribution of the \texttt{GEI} node, with the probability associated with its highest level increasing from 0.29 to 0.42. As noted for the domains, the obtained GEI distribution does not merely reflect the indicator’s computational structure but instead embodies the complex information flow that propagates across ingredients, domains, and extra-variables. These results indicate that the North-East, which begins from relatively good equality conditions, can still register additional gains when such policies are applied.

When the same scenario is assessed in the South macro-region, it is important to account for the fact that southern regions face substantially more severe gender inequalities than the North-East, as indicated by the baseline probability distributions in Table \ref{tab:extra_scen}. Specifically, for every domain, provinces in the South of Italy exhibit substantially higher probabilities of being at the lowest level of gender equality. In the \texttt{Money Domain}, for example, the probability of being in the lowest state is 0.77, approximately five times higher than the value recorded for the North-East (0.15). This province-level differential cascades consistently onto the GEI distribution, with the southern region exhibiting probabilities of 0.87 and 0.07 corresponding to the lowest and highest levels of equality, respectively, underscoring the pronounced levels of gender inequality in this macro-region.

As shown in Table \ref{tab:extra_scen}, simulating a policy in the southern provinces that sets \texttt{Work segregation equality} and \texttt{Employment rate} to their highest levels is associated with improved equality conditions across most domains, as also observed in the North-East. However, while the latter already exhibits relatively higher equality levels after the policy implementation, the southern regions show the largest increase. Moreover, the \texttt{Power Domain} distribution, which was largely unaffected by the policy in the North-East, shows a substantial shift toward its right tail in the southern provinces, with the probability of its lowest state decreasing from 0.42 to 0.12 and that of its highest state increasing from 0.12 to 0.43. This differential response may reflect the fact that increases in employment and reductions in work segregation have stronger spillover effects in territories where women’s economic participation and representation in decision-making bodies are initially low. In the South, such improvements can translate into a larger presence of women-led enterprises and greater involvement in local administrations, thereby shifting the Power Domain upward, whereas in the North-East, where these indicators are already relatively high, the scope for further gains is more limited. The marked improvement across domains is also reflected in the GEI distribution, with the probability of the lowest equality level dropping from 0.87 to 0.49, and the highest equality level rising from 0.07 to 0.35.

Taken together, the analysis reveals two markedly different regional dynamics. In the Italian North-East, the initial equality conditions are already relatively favorable, and the simulated policy further enhances these levels, producing a consistently high equality profile across domains. In contrast, the South  of Italy begins from substantially lower starting values, and although the policy does not bring equality levels close to those observed in the North-East, it nevertheless generates meaningful improvements, markedly advancing the region’s overall equality conditions. These differentiated outcomes highlight the importance of conducting analyses that incorporate the socio-economic context and the interactions between GEI ingredients and broader territorial characteristics. Such contextualization is made possible by the OOBN framework, which embeds the GEI computation within a region’s specific socio-economic environment. This emerging evidence underscores the value of employing Object-Oriented Bayesian Networks for GEI modeling, as they enable tailored and effective ex-ante policy evaluation. By capturing the multivariate relational structure linking extra-variables, ingredients, and domains, the OOBN framework proves to be a powerful tool for informing and guiding evidence-based policy design to address gender inequalities.

\section{Concluding Remarks }\label{sec5}

The novel methodology proposed in this paper leverages Object-Oriented Bayesian Networks (OOBNs) to enhance the use of the Gender Equality Index (GEI). By moving beyond a single Bayesian Network (BN)  model, the proposed approach reflects and preserves the hierarchical architecture of the GEI, addresses its limitations as a composite indicator, and delivers enriched analytical outputs. In doing so, it improves the understanding, assessment and monitoring of gender equality.

Through a case study based on Italian official statistics, the paper illustrates several advantages offered by the OOBN framework when applied to the GEI. In particular:
\begin{enumerate}[(a) ]
\item OOBNs can naturally accommodate prior knowledge and additional variables, allowing the assessment of gender equality to be better aligned with the specific demographic, social, and economic context of a country;
\item As a comprehensive multivariate statistical learning model, OOBNs enable an advanced quantitative assessment of gender equality, shifting the focus from the mere computation of a single GEI score to the modelling of the full system of conditional (in)dependencies among all measured variables;
\item Through their graphical component, OOBNs provide a qualitative and interpretable representation of the dimensions used to reflect gender equality, together with their associations, interactions, and intersectional influences. This supports the identification of underlying mechanisms of gender-based inequality, discrimination, and exclusion, and offers a strategic framework for programme design and management aimed at achieving effective gender equality outcomes;
\item By adding a probabilistic dimension, OOBNs empower the GEI score with the ability to simulate policy scenarios. This feature directly supports Gender Impact Assessment and evidence-based decision-making through ex-ante evaluation of gender equality policies, actions, and resource allocations.
\end{enumerate}

The hierarchical and modular architecture of the OOBN supports multi-level decision-making aligned with the layers of the hypernetwork (\ref{fig:Fig12}). Domain-specific interventions can be evaluated within the corresponding BN instances (\ref{fig:CombinedCaption}), while their combined effects are assessed at the top level of the model.
More generally, the results of this study confirm that assessing gender equality and its policy impacts requires not only information on gender differences, but a deeper understanding of the mechanisms through which such differences translate into inequalities, discrimination, and exclusion. By making these mechanisms explicit, the OOBN framework provides an operational statistical tool to move from descriptive monitoring toward analytical explanation and policy-oriented evaluation.

The proposed methodology also presents limitations. As with any data-driven statistical model, OOBNs are sensitive to the availability, quality, and granularity of data. Estimated probability distributions and scenario-based insights are necessarily constrained by the information content of the data used for model learning. These limitations emerged clearly in the Italian case study (see Section~\ref{sec4}), most notably in the lack of data for variables associated with the Time domain of the GEI, which required the estimation of an OOBN based on five domains instead of six. Increasing demands for analytical granularity further exacerbated data sparsity, necessitating the use of proxy variables across several GEI ingredients. To mitigate these risks and preserve the effectiveness of the proposed approach, the use of official, harmonized, and highly reliable data sources is therefore essential.

At the same time, the modular and object-oriented nature of the proposed framework makes it highly transferable and scalable. The OOBN approach can be adapted to different national and sub-national contexts, to specific population groups, and to alternative gender composite indicators, as well as to composite indicators in other policy domains. By combining data-driven structural learning with a modelization that preserves hierarchical indicator architectures, OOBNs provide a general and flexible methodological framework that advances intersectional analysis and strengthens the role of composite indicators as decision-support tools.

In this perspective, the use of OOBNs to empower the GEI responds directly to the need, highlighted in the Introduction (Section \ref{sec1}), for state-of-the-art statistical tools capable of complementing composite indicators. Beyond benchmarking and monitoring, the proposed approach enables explanatory and predictive analyses that support Gender Impact Assessment and more informed, transparent, and effective gender-sensitive policymaking.

\bibliographystyle{abbrvnat}
\bibliography{reference}

\end{document}